\documentclass[12pt,preprint]{aastex}
\usepackage{amssymb,amsmath}
\usepackage{color,hyperref}
\definecolor{linkcolor}{rgb}{0,0,0.25}
\hypersetup{
  colorlinks=true,        
  linkcolor=linkcolor,    
  citecolor=linkcolor,    
  filecolor=linkcolor,    
  urlcolor=linkcolor      
}
\newcounter{address}
\newcommand{\etal}{et al.}
\newcommand{\dd}{\mathrm{d}}
\newcommand{\eg}{e.g.}
\newcommand{\eqnname}{equation}
\newcommand{\Eqnname}{Equation}

\newcommand{\sectionname}{$\mathsection$}

\newcommand{\kms}{\,\mbox{km s}^{-1}}
\newcommand{\kpc}{\,\mbox{kpc}}
\newcommand{\pc}{\,\mbox{pc}}
\newcommand{\msun}{\,M_\odot}

\newcommand{\mb}{MB12}

\begin{document}

\title{On the local dark matter density}
\author{Jo~Bovy\altaffilmark{1} and Scott Tremaine}
\affil{Institute for Advanced Study, Einstein Drive, Princeton, NJ
  08540, USA}
\altaffiltext{\theaddress}{\stepcounter{address} Hubble
  fellow}

\begin{abstract} 
  An analysis of the kinematics of 412 stars at 1--$4\kpc$ from the
  Galactic mid-plane by Moni Bidin \etal\ (2012b) has claimed to derive
  a local density of dark matter that is an order of magnitude below
  standard expectations. 
  We show that this result is incorrect and that it arises from the
  assumption that the mean azimuthal velocity of the stellar tracers
  is independent of Galactocentric radius at all heights.  We
  substitute the assumption, supported by data, that the circular
  speed is independent of radius in the mid-plane.  We demonstrate
  that the assumption of constant mean azimuthal velocity is
  implausible by showing that it requires the circular velocity to
  drop more steeply than allowed by any plausible mass model, with or
  without dark matter, at large heights above the mid-plane. Using the
  approximation that the circular velocity curve is flat in the
  mid-plane, we find that the data imply a local dark-matter density
  of $0.008\pm0.003\msun\pc^{-3}= 0.3\pm0.1\,\mbox{GeV cm}^{-3}$,
  fully consistent with standard estimates of this quantity.
  This is the most robust direct measurement of the local dark-matter
  density to date.
\end{abstract}

\keywords{
        Galaxy: disk
        ---
        Galaxy: fundamental parameters
        ---
        Galaxy: halo
        ---
        Galaxy: kinematics and dynamics
        ---
        Galaxy: solar neighborhood
        ---
        Galaxy: structure
}

\section{Introduction}

The observed flatness of the Milky Way's circular-velocity curve at
Galactocentric distances larger than 20 kpc \citep[\eg,][]{Xue08a}
shows that the visible Galactic disk is embedded in a massive dark
halo. The disk is composed of gas and stars (baryons), while the dark
halo is believed to be composed of non-baryonic matter of unknown
nature. Despite the dominance of the dark halo in the outer parts of
the Milky Way, it
remains unclear from direct measurements whether there is any need for
a substantial amount of dark matter to explain the circular-velocity
curve interior to the solar radius, $R_0 = 8\kpc$. Quantitatively, the
fraction of the total radial force at the solar radius that is due to
the disk could be as high as 90\% \citep{Sackett97a}. A promising
avenue for constraining the local density of dark matter is through a
determination of the dependence of the gravitational potential on
distance above the mid-plane of the disk (``height''), from measuring
the kinematics of stars
\citep[\eg,][]{Kapteyn22a,Oort32a,Bahcall84a}. However, a major
obstacle is that the uncertainty in the amount of baryonic
matter in the disk makes it hard to determine the relative
contributions from dark and baryonic matter to the density near the
mid-plane. Studies limited to heights $\lesssim 150$ pc are consistent
with no dark matter near the Sun, but they cannot exclude the amount
of dark matter expected for a standard dark-matter halo ($\approx
0.01\, M_\odot\pc^{-3}= 0.38\,\mbox{GeV cm}^{-3}$;
\citealt{Creze98a,Holmberg00a}).

The contributions from baryonic and dark matter can be disentangled by
measuring the gravitational potential out to larger heights. At
heights of several times the disk thickness, the dark halo and the
baryonic disk contributions to the potential have a different vertical
dependence \citep[\eg,][]{Kuijken89a,Garbari11a}. In particular, most
of the disk mass lies below $\sim1.5\kpc$, so above this height the
disk contribution to the integrated surface density
$\Sigma(Z)\equiv\int_{-Z}^Z\dd z\,\rho(z)$ is roughly constant with
height and the disk potential varies as $\Phi \propto |Z|$, while the
dark halo contributes $\Sigma(Z)\propto |Z|$ and $\Phi(Z)\propto Z^2$.
Thus, any measured increase in the surface density at $|Z|\gtrsim
2\kpc$ must be due to the dark halo, and the expected increase is
$\sim 20\msun\,\mbox{pc}^{-2}$ for each kpc (for the standard value of
the local dark matter density of $0.01\msun\,\mbox{pc}^{-3}$; see
above). Determinations of the surface density at $\sim\!1\kpc$ from
the plane typically find values of $\Sigma(1\kpc)=70\mbox{ to }
80\,M_\odot\pc^{-2}$ 
\citep{Kuijken91a,Siebert03a,Holmberg04a} while the baryonic
contribution is estimated to be around $50\mbox{ to }\,60 M_\odot\pc^{-2}$
\citep{Holmberg04a,Bovy12a}. Thus, these studies are consistent with
the expected dark-matter density, although they do not go to large
enough heights to detect the dark matter unambiguously.

A recent study of the kinematics of stars with heights $1.5 \lesssim
|Z| \lesssim 4\kpc$ by \citet[][\mb\ hereafter]{MoniBidin12b} claims
to have measured the surface density at these heights in a
model-independent manner and found it to be constant, such that the
local density of dark matter must be smaller than $10^{-3} M_\odot\pc^{-3}$ ($< 0.04\,\mbox{GeV cm}^{-3}$). If true, this would imply that
dark matter is less abundant in the solar neighborhood than expected,
by at least an order of magnitude. This would also shift experimental
limits on the cross-section for elastic scattering between dark-matter
particles and baryons by an order of magnitude, although this is far
less than the uncertainty in the predicted cross-section.

In this paper, we show that the analysis used by \mb\ is flawed. The
main error is that they assume that the \emph{mean azimuthal} (or
{\emph rotational}) \emph{velocity} $\bar{V}$ of their tracer
population is independent of Galactocentric cylindrical radius $R$ at
all heights (i.e., $\bar{V}(R,Z)=\bar{V}(Z)$). This assumption is not
supported by the data, which instead imply only that the circular
speed $V_c$ is independent of radius in the mid-plane
\citep[e.g.,][]{GKT79,Feast97a}. In the solar neighborhood, the
circular speed is larger by $\gtrsim 35\,\kms$ than the mean azimuthal
velocity for the warm tracer population used by \mb, a phenomenon
known as asymmetric drift. The asymmetric drift \emph{is} expected to
vary with $R$---although this variation cannot be measured for the
sample of \mb\ as the data do not span a large enough range in
$R$---so the assumptions that $\bar{V}$ and $V_c$ are independent of
radius are not compatible. In the absence of a measurement of the
$(R,Z)$ dependence of $\bar{V}$ for the tracers, we show that the
assumption of an $R$-independent $\bar{V}$ at all heights $Z$ is
highly implausible. By using instead the data-driven assumption that
the circular speed is independent of radius at $Z = 0$, we demonstrate
that the measurements and analysis of \mb\ are fully consistent with
the standard estimate of the local dark matter density, approximately
$0.01\, M_\odot\pc^{-3}$, and indeed provide the best available direct
measurement of this quantity.

The outline of this paper is as follows. In
\sectionname~\ref{sec:vphivcirc} we discuss the assumption made by
\mb\ that the mean azimuthal velocity of the tracer population is
independent of $R$. In \sectionname~\ref{sec:poisson}, we derive the
surface density as a function of $Z$ using the assumption that
the circular-velocity curve is flat in the mid-plane and we discuss
the effect of relaxing this approximation. In
\sectionname~\ref{sec:reanalysis}, we calculate the surface
density using the data of \mb\ but using the approximation of constant
circular speed and show that this leads to a surface density at
$1.5 < |Z| < 4\kpc$ that increases with height at the rate expected
for a dark halo. Our conclusions are in
\sectionname~\ref{sec:conclusion}. 

We follow \mb\ in defining $U$, $V$, and $W$ as the radial, azimuthal,
and vertical velocities in cylindrical coordinates and the inertial
Galactocentric reference frame. Note that this definition is
non-standard, since (i) $U$, $V$ and $W$ are normally defined with
respect to the Local Standard of Rest; (ii) $U$ is typically the
velocity toward the Galactic center; (iii) our conventions require
either that the coordinate system is left-handed or that the positive
$z$-axis points to the South Galactic Pole. However, for the purpose
of this paper these distinctions do not matter. We use $R_0 = 8\kpc$,
as do \mb, and local circular speed $V_c = 220\,\kms$ throughout this
paper, assumptions which are consistent with the latest measurements
\citep{Bovy09a}. We abbreviate the surface density at the solar radius
as $\Sigma(Z) \equiv \Sigma(R_0,Z) = \int_{-Z}^Z \dd z\, \rho(R_0,z)$.

\section{Mean azimuthal velocity versus circular velocity}\label{sec:vphivcirc}

The analysis of \mb\ uses the Poisson equation to
calculate the surface density as a function of height $Z$ at the solar
radius $R_0$ using the radial force $F_R$ and the vertical force $F_Z$
(eq.\ \ref{eq:fz}). To estimate $F_R$,
\mb\ make use of the radial Jeans equation
\begin{equation}\label{eq:radialjeans}
  F_R(R,Z) = -\frac{\partial \Phi(R,Z)}{\partial R} = \frac{1}{\nu}\,\frac{\partial \left(\nu\sigma_U^2\right)}{\partial R} +\frac{1}{\nu}\,\frac{\partial \left(\nu\sigma^2_{UW}\right)}{\partial Z} + \frac{\sigma_U^2 - \sigma_V^2 - \bar{V}^2}{R}\,,
\end{equation}
where $\Phi$ is the gravitational potential, $\nu$ is the
tracer-density profile, $\sigma^2_U$ and $\sigma^2_V$ are the radial
and azimuthal velocity dispersions squared, $\sigma^2_{UW}$ is the
off-diagonal radial--vertical entry of the dispersion-squared matrix,
and $\bar{V}$ is the mean azimuthal velocity; all of these quantities
are functions of $R$ and $Z$.  We have assumed a steady state such
that time derivatives vanish and the mean radial motion $\bar{U}$ is
zero. In simplifying \eqnname~(\ref{eq:radialjeans}), \mb\ use their
assumption VIII: ``The rotation curve is locally flat in the volume
under study'', which \mb\ express as
\begin{equation}
\frac{\partial \bar{V}(R_0,Z)}{\partial R} = 0\,.
\end{equation}
\citet{MoniBidin10a} made the same assumption, although this was not
explicitly stated. We show in this section that this is an
unreasonable assumption.

The mean azimuthal velocity of a population of stars differs from the
circular velocity due to the asymmetric drift. This offset arises
because both the density of stars and the velocity dispersion
typically decline with radius. This
means that more stars with guiding centers at $R < R_0$ are passing
through the solar neighborhood than stars with guiding centers $R >
R_0$; the former are on the outer parts of their orbits, where their
azimuthal velocity is less than the circular
velocity. \Eqnname~(\ref{eq:radialjeans}) is typically used to
estimate the asymmetric drift \citep{binneytremaine}: since $V_c^2 =
R\,\partial \Phi / \partial R$ we find
\begin{equation}
  V_c^2 - \bar{V}^2 = \sigma_V^2 -\sigma_U^2\,\left[1 + \frac{\partial
      \ln\left(\nu\sigma_U^2\right)}{\partial \ln R} +
    \frac{1}{\nu}\,\frac{R}{\sigma_U^2}\,\frac{\partial\left( \nu
        \sigma^2_{UW}\right)}{\partial Z}\right]\,.
\label{eq:adrift}
\end{equation}

Now assume that (i) the dispersions-squared
($\sigma_U^2,\sigma_V^2,\sigma^2_{UW}$) decline exponentially
with radius with scale length $h_\sigma$; (ii) the tracer density is
an exponential function of radius and height with scale lengths $h_R$
and $h_Z$, that is, $\nu(R,Z)\propto \exp(-R/h_R-|Z|/h_Z)$. The second
assumption is only accurate at heights above a few hundred pc; closer to
the mid-plane the exponential form is not accurate but this is not a
concern since we are interested in the region $|Z|\gtrsim
1\kpc$. These assumptions were also made by \mb\ and in addition they
assumed that $h_\sigma=h_R$. Equation (\ref{eq:adrift}) then becomes 
\begin{equation}\label{eq:barv}
  V_c^2 - \bar{V}^2 = \sigma_V^2+\sigma_U^2\,\left[ R\left(\frac{1}{h_R}+\frac{1}{h_\sigma}\right) -1\right]
   + \frac{R}{h_Z}\,\sigma^2_{UW} -R\,\frac{\partial
      \sigma^2_{UW}}{\partial Z}. 
\end{equation}
To evaluate this, we use the expressions for the moments of the velocity distribution from \citet{MoniBidin12a}
\begin{align}
  \sigma_U(R_0,Z) &= (82.9 \pm 3.2) + (6.3 \pm 1.1)\cdot(|Z|/\mathrm{kpc}-2.5)\ \mathrm{km\ s}^{-1}\\
  \sigma_V(R_0,Z) &= (62.2 \pm 3.1) + (4.1 \pm 1.0)\cdot(|Z|/\mathrm{kpc}-2.5)\ \mathrm{km\ s}^{-1}\\
  \sigma_W(R_0,Z) &= (40.6 \pm 0.8) + (2.7 \pm 0.3)\cdot(|Z|/\mathrm{kpc}-2.5)\ \mathrm{km\ s}^{-1}
\end{align}
 and that for $\sigma^2_{UW}$ from \mb
\begin{equation}
  \sigma^2_{UW}(R_0,Z) = (1522 \pm 100) + (366 \pm 30)\cdot(|Z|/\mathrm{kpc} - 2.5)\ \mathrm{km}^2\ \mathrm{s}^{-2}\,.
\end{equation}
We also take $h_z=0.9\kpc$ and $h_R = h_\sigma = 3.8\kpc$ as in \mb,
although our own work suggests somewhat smaller values $h_R=2\kpc$ and
$h_\sigma=3.5\kpc$ \citep{Bovy12c}. We find that the solution to
equation (\ref{eq:barv}) can be fit by the formula
\begin{equation}
  V_c^2-\bar{V}^2 =
  (191\kms)^2\,\left[1+0.19\,\left(|Z|/\mathrm{kpc}-2.5\right)\right]\,,
\label{eq:adriftb}
\end{equation}
with an rms error of less than 2\% for $|Z|=1$--$4\kpc$. Equation
(\ref{eq:adriftb}) implies $V_c-\bar{V}(1.5\ \mathrm{kpc}) = 80$ km
s$^{-1}$ and $V_c-\bar{V}(3.5\ \mathrm{kpc}) = 150\kms$, in
good agreement with the corresponding measured values in
\citet{MoniBidin12a}, $70\pm 13\kms$ and $130\pm 16\kms$.

We can now ask what \eqnname~(\ref{eq:barv}) predicts for the radial
variation of $\bar{V}$. Taking the radial derivative of
\eqnname~(\ref{eq:barv}), we find that
\begin{align}\label{eq:barvderiv}
  2\,V_c\,\frac{\partial V_c}{\partial R} - 2\,\bar{V}\,\frac{\partial \bar{V}}{\partial R}
  &= -\frac{(V_c^2 - \bar{V}^2)}{h_\sigma} +
  \sigma_U^2\,\left(\frac{1}{h_R} + \frac{1}{h_\sigma}\right) +
  \frac{1}{h_Z}\,\sigma^2_{UW}-\frac{\partial \sigma^2_{UW}}{\partial Z}\,\nonumber \\
&=(V_c^2 - \bar{V}^2)\left(\frac{1}{R}-\frac{1}{h_\sigma}\right)+\frac{\sigma_U^2}{R}-\frac{\sigma_V^2}{R}\,.
\end{align}
We can then estimate $\partial \bar{V}/\partial R$
for a flat circular-speed curve, by setting $\partial V_c / \partial R
= 0$:
\begin{equation}
  \bar{V}\,\frac{\partial \bar{V}}{\partial R} = 110\kms\times21\kms\kpc^{-1}\,\left[1+0.2\,\left(|Z|/\mathrm{kpc}-2.5\right)\right]\,,
\end{equation}
with an rms error of less than 1\% for $|Z|=1$--$4\kpc$.  In this
equation, $110\kms$ and $21\kms\kpc^{-1}$ are $\bar{V}$ and
$\partial\bar{V}/\partial R$ at $|Z|=2.5\kpc$. We find that $\partial
\bar{V}/\partial R$ is $7\kms\kpc^{-1}$ at $Z=0$, growing to
$11\kms\kpc^{-1}$ at $|Z|=1\kpc$ and $40\kms\kpc^{-1}$ at
$|Z|=3.5\kpc$. If we use $h_\sigma=3.5\kpc$ \citep{Bovy12c}, the
gradients are larger by about 20\%. 

For comparison, \mb\ estimate that a gradient $\partial
\bar{V}/\partial R=17\kms\kpc^{-1}$ is needed to make their analysis
consistent with the expected amount of dark matter, which is close to
our estimate of $21\kms\kpc^{-1}$. \mb\ dismiss this
possibility, apparently because they confuse constraints on $\partial
V_c / \partial R$ with constraints on $\partial \bar{V} / \partial R$,
as there are, in fact, no direct observational constraints on
$\partial \bar{V} / \partial R$.

While $\partial \bar{V} / \partial R$ is not measured for the \mb\
sample, constraints on this gradient do exist for a
slightly more metal-rich sample of stars at moderate distances from
the mid-plane, which we can use to show that our estimate of
$\partial\bar{V}/\partial R$ using \eqnname~(\ref{eq:barvderiv})
agrees with observations. \citet{CasettiDinescu11} study a sample of
metal-rich red clump stars at $1.0\kpc \leq |Z|\leq 2.5\kpc$, selected
using $0.6 \leq J-K_s \leq 0.7$, while the \mb\ data were selected
using $0.7 \leq J-K_s \leq 1.1$ to obtain a lower-metallicity,
larger-distance sample. \citet{CasettiDinescu11} report that
$(\sigma_U,\sigma_V,\bar{V}) = (60,42,180)\kms$ at $|Z| \approx
1\kpc$, such that \eqnname~(\ref{eq:barvderiv}) assuming a flat
circular-speed curve predicts that $\partial\bar{V}/\partial R =
6.5\kms\kpc^{-1}$; similarly, they find $(\sigma_U,\sigma_V,\bar{V}) =
(80,60,150)\kms$ at $|Z| \approx 1.5\kpc$ and
$(\sigma_U,\sigma_V,\bar{V}) = (90,65,140)\kms$ at $|Z| \approx
2\kpc$, such that we predict that $\partial\bar{V}/\partial R =
13\kms\kpc^{-1}$ and $15\kms\kpc$, respectively. These predictions are
in good agreement with the measurements of \citet{CasettiDinescu11},
who find $6.0\pm1.5\kms\kpc^{-1}$, $13.0\pm4.5\kms\kpc^{-1}$, and
$12.0\pm8.1\kms\kpc^{-1}$ at these $|Z|$.

The discussion so far in this section has assumed that $V_c$ is
constant with $Z$ and $R$, whereas observations only show that $V_c$
is independent of $R$ at $Z=0$. The circular velocity declines with
$|Z|$ for any reasonable local mass distribution, as the following
argument shows.  As $V_c^2 = -R\,F_R$, we can write
\begin{equation}
  \frac{\partial V_c^2}{\partial Z} = -\frac{\partial}{\partial
    Z}\,\left(R\,F_R\right)
= -R\frac{\partial F_Z}{\partial R}\,,
\end{equation}
where we have used the fact that $F_R$ and $F_Z$ are both derivatives
of the potential $\Phi$, such that 
\begin{equation}\begin{split}
\frac{\partial F_R}{\partial Z} &=
-\frac{\partial}{\partial Z} \left(\frac{\partial \Phi}{\partial R}\right)
=-\frac{\partial}{\partial R} \left(\frac{\partial \Phi}{\partial Z}\right)
= \frac{\partial F_Z}{\partial R}\,.
\end{split}\end{equation}
As shown below in \sectionname~\ref{sec:poisson}, to a good
approximation $F_Z = -2\pi G \Sigma(R,Z)$ (for $Z > 0$), where
$\Sigma(R,Z)$ is the surface density. Therefore
\begin{equation}\label{eq:dvcdz0}
  2\,V_c\,\frac{\partial V_c}{\partial Z} \simeq 2\pi GR\,\frac{\partial
    \Sigma(R,Z)}{\partial R}=-2\pi G \frac{R}{h_\Sigma}\,\Sigma(R,Z)\,,
\end{equation}
assuming that $\Sigma(R,Z)$ declines exponentially with radius with
scale length $h_\Sigma$, which we assume is equal to $3.5\kpc$
\citep{Bovy12c}. This means that near the mid-plane, where the density
is approximately $0.1\msun\pc^{-3}$, 
\begin{equation}\label{eq:dvcdz1}
  \frac{\partial V_c}{\partial Z} \simeq -2.8\, \mathrm{km\ s}^{-1}\ \mathrm{kpc}^{-1}\, \left(\frac{Z}{100\,\mathrm{pc}}\right)\,\qquad |Z| \lesssim h_Z\,,
\end{equation}
while for $|Z| > 1.5\kpc$, where
$\Sigma(R_0,Z) \simeq 50\, M_\odot$ pc$^{-2}$
\begin{equation}\label{eq:dvcdz2}
  \frac{\partial V_c}{\partial Z} \simeq -7\, \mathrm{km\ s}^{-1}\ \mathrm{kpc}^{-1}\,\qquad |Z| \gg h_Z\,.
\end{equation}
Taken together, these results imply that $V_c$ likely does not
decrease by more than $30\kms$ out to $|Z| = 4\kpc$. Such a
decrease does not change the conclusions of this section.

A different way of illustrating the inconsistency of \mb's assumptions
is to ask what the radial behavior of $V_c$ has to be such that
$\partial \bar{V} / \partial R = 0$. The necessary $\partial V_c /
\partial R$ decreases from approximately $-5\kms\kpc^{-1}$
at $Z=0$,
which is consistent with local measurements of the slope of the
circular-speed curve, to $-14\kms\kpc^{-1}$ at $|Z| = 4\kpc$; for
$h_\sigma=3.5\kpc$, the gradients are larger in absolute value by
20\%.  Such a steep drop of the circular velocity with $R$ is about
the value of a Keplerian drop-off, $-\frac{1}{2}V_c/R_0 = -14\kms\kpc^{-1}$. Such steep gradients are inconsistent with observational
evidence: (i) in the limiting case where there is no dark-matter halo
and all the mass is in an exponential disk (with parameters given in
\figurename~\ref{fig:approx}), $|\partial V_c/\partial R|$ is only
$3\kms\kpc^{-1}$ at $|Z|=4\kpc$, and if a dark halo is present the
value is even smaller; (ii) we show below in
\sectionname~\ref{sec:poisson} that $\partial V_c / \partial R$
\emph{increases} as $|Z|$ grows, whereas maintaining constant
$\bar{V}$ requires that $\partial V_c / \partial R$ \emph{decreases}
with $|Z|$.

\section{Poisson equation at large heights}\label{sec:poisson}

Having shown that the assumption of a radially constant $\bar{V}$ is
suspect, we now ask what the \mb\ data do have to say about the
surface density at large height $|Z|$. Following \mb, we start from the
Poisson equation in cylindrical coordinates,
\begin{equation}\label{eq:poisson}
\Sigma(R,Z) = -\frac{1}{2\pi G}\,\left[\int_{0}^Z \dd z\,\frac{1}{R} \,\frac{\partial(RF_R)}{\partial R} + F_Z(R,Z)\right]\,,
\end{equation}
where we have assumed symmetry around the $Z = 0$ plane. $F_Z(R,Z)$ is
obtained from the steady-state vertical Jeans equation
\begin{equation}
  F_Z(R,Z) = -\frac{\partial \Phi(R,Z)}{\partial Z} =
  \frac{1}{\nu}\,\frac{\partial (\nu \sigma_W^2)}{\partial Z} +
  \frac{1}{R\nu}\,\frac{\partial \left(R\nu\sigma^2_{UW}\right)}{\partial
    R}\,.
\label{eq:fz}
\end{equation}
One can then proceed
\citep[\eg,][]{Kuijken89a} by approximating the integrand in the first
term in square brackets in \eqnname~(\ref{eq:poisson}) by its value in
the plane. This is zero for a flat circular-speed curve, as $R F_R
=-V_c^2$. \citet{Kuijken89a} show that this approximation is good to a
few percent at $|Z| \lesssim 1.5\kpc$ for any plausible mass
distribution. We now revisit this question at larger heights.

\begin{figure}[htbp]
\includegraphics[width=0.8\textwidth,clip=]{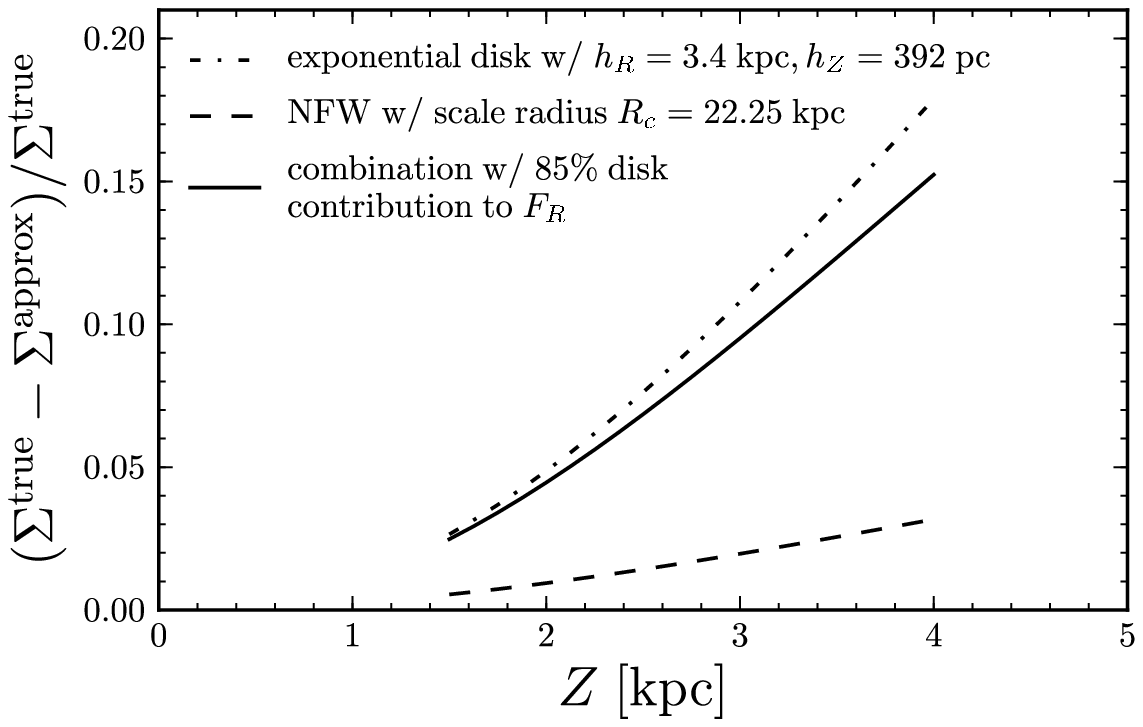}
\caption{Fractional difference between the true surface density
  $\Sigma^{\mathrm{true}}(R_0,Z)$ and that obtained by approximating
  the integrand in \eqnname~(\ref{eq:poisson}) by its value in the
  plane, $\Sigma^{\mathrm{approx}}(R_0,Z)$. Shown are an exponential
  disk, a spherical NFW halo, and a combination of the two
  that has a circular-speed curve that is flat near $R_0 = 8$
  kpc.}\label{fig:approx}
\end{figure}

\figurename~\ref{fig:approx} shows the error introduced by
approximating $\partial\left(RF_R\right)/\partial R$ by its value
at $Z = 0$, for three different mass distributions: an exponential
disk with parameters that are representative of the Milky Way's disk
\citep{Bovy12a,Bovy12b}, a Navarro-Frenk-White (NFW;
\citealt{Navarro97a}) halo, and a combination of the two in which the
disk provides 85\,\% of the radial force at $R_0$ (this yields a flat
circular-speed curve at $R_0$). This figure shows the fractional
difference between the true surface density and that calculated from
equation (\ref{eq:poisson}) by approximating $\partial\left( R F_R
\right)/ \partial R$ by its value in the plane. The errors introduced
by this approximation can be as large as 15\,\% at 4 kpc above the
plane. Thus a systematic uncertainty of 10 to 20\,\% in
$\Sigma(4\ \mathrm{kpc})$ is unavoidable without fully modeling the
potential.

It is, however, possible to show that neglecting the $Z$-dependence of
$\partial V_c / \partial R$ almost always leads to an \emph{underestimate}
of the surface density, so this approximation gives a robust
lower limit on the surface density. We start by writing the $Z$
derivative of the integrand in equation (\ref{eq:poisson}) as
\begin{equation}
\frac{\partial}{\partial Z} \left(\frac{1}{R}\,\frac{\partial(R
  F_R)}{\partial R}\right) = \frac{\partial}{\partial Z}
\left(\frac{F_R}{R} + \frac{\partial F_R}{\partial R}\right)\,.
\end{equation}
If we again use the fact that $\partial F_R / \partial Z =
\partial F_Z / \partial R$, we find that
\begin{equation}\label{eq:test}
\frac{\partial}{\partial Z} \left(\frac{1}{R}\,\frac{\partial(R
  F_R)}{\partial R}\right) = \frac{1}{R}\,\frac{\partial F_Z}{\partial R}
+ \frac{\partial^2 F_Z}{\partial R^2}\,.
\end{equation}
If the first term in square brackets in equation (\ref{eq:poisson}) is
neglected, we have $F_Z = -2\pi G \Sigma(R,Z)$
(at $Z > 0$). This approximation is quite accurate if there is
substantial mass in a thin disk and $|Z|$ is
small, or if the circular-speed curve is nearly flat, since then
$RF_R$ is almost independent of radius (see
\figurename~\ref{fig:approx}). If we then drop this term and approximate the radial derivatives then as those
of the exponential disk, equation (\ref{eq:test}) becomes 
\begin{equation}\label{eq:zderivapprox}
\frac{\partial}{\partial Z} \left(\frac{1}{R}\,\frac{\partial(R
  F_R)}{\partial R}\right) \simeq \frac{2\pi G}{R\,h_\Sigma} \,\left(1 -
\frac{R}{h_\Sigma}\right)\,\Sigma(R,Z)\,.
\end{equation}
Since $R_0/h_\Sigma$ is 2 or more in
any reasonable model for the Milky Way, the right side of this
equation is negative, which in turn implies that the integrand in
equation (\ref{eq:poisson}) decreases with increasing $|Z|$. Thus
replacing $F_R$ by its value in the mid-plane leads to an 
\emph{underestimate} of the surface density, and dropping the integral
in equation (\ref{eq:poisson}) when the circular-speed curve is
locally flat also leads to an underestimate of the surface density. 

We can further work out \eqnname~(\ref{eq:zderivapprox}) to find the
change in the radial slope of the circular-speed curve with height
\begin{equation}
  \frac{\partial}{\partial Z} \left(\frac{\partial V_c}{\partial
    R}\right) \simeq \frac{\pi
    G}{h_\Sigma\,V_c}\,\left(\frac{R}{h_\Sigma}-1\right) \Sigma(R,Z)\,,
\end{equation}
a result that can also be derived by differentiating equation
(\ref{eq:dvcdz0}). Using the same reasoning that led to \eqnname
s~(\ref{eq:dvcdz1}--\ref{eq:dvcdz2}), we find
\begin{equation}
  \frac{\partial}{\partial Z} \left(\frac{\partial V_c}{\partial
    R}\right) \simeq 0.45\,\mathrm{km\ s}^{-1}\ \mathrm{kpc}^{-2}\,
  \left(\frac{Z}{100\,\mathrm{pc}}\right)\,,\,\qquad |Z| \lesssim h_Z\,,
\end{equation}
and
\begin{equation}
  \frac{\partial}{\partial Z} \left(\frac{\partial V_c}{\partial
    R}\right) \simeq
  1.1\,\mathrm{km\ s}^{-1}\ \mathrm{kpc}^{-2}\,,\,\qquad |Z| \gg
  h_Z\,.
\end{equation}
Thus, the radial gradient of the circular speed remains close to its
value in the plane throughout the region $|Z|<4\kpc$ that we are
investigating.

\section{Moni Bidin \etal\ (2012) revisited}\label{sec:reanalysis}

We have shown that approximating the integrand  in \eqnname~(\ref{eq:poisson}) as constant with $Z$ leads to a robust lower
limit on the surface density as a function of $Z$. Armed with this result, we can revisit
the \mb\ analysis. Since the circular-speed curve is flat in the
mid-plane, the integrand is zero in the mid-plane, so the
approximation that the integrand is constant implies that the integral
can be neglected. Substituting equation (\ref{eq:fz}) into equation
(\ref{eq:poisson}) we find 
\begin{equation}\label{eq:correct}
  \Sigma(Z) = -\frac{1}{2\pi G}\,\left[-\frac{1}{h_Z}\,\sigma_W^2 + \frac{\partial \sigma_W^2}{\partial Z} + \sigma^2_{UW}\,\left(\frac{1}{R} - \frac{1}{h_R} - \frac{1}{h_\sigma}\right)\right]\,,
\end{equation}
where we distinguish between the radial scale length $h_R$ of the
tracer population and the radial scale length $h_\sigma$ of the
dispersion-squared. \mb\ assumed $h_R=h_\sigma=3.8\kpc$, but recent
direct measurements of the thicker populations of stars in the Milky
Way have shown that their radial scale length is much shorter than
that of the dominant, thinner components, such that $h_R = 2\kpc$ is
more accurate; we also prefer a shorter scale length for the
dispersions, $h_\sigma=3.5\kpc$ \citep{Bovy12b,Bovy12c}. We will
nevertheless show results mostly for the values $h_R = h_\sigma =
3.8\kpc$ assumed by \mb. We also follow \mb\ in assuming $h_Z = 900$
pc (see below). Note that the surface density as calculated using
\eqnname~(\ref{eq:correct}) does \emph{not} depend on the value of the
local circular speed $V_c$, and our estimate of the local dark
matter density is therefore not affected by the uncertainty in $V_c$.

\begin{figure}[htbp]
\includegraphics[width=0.75\textwidth,clip=]{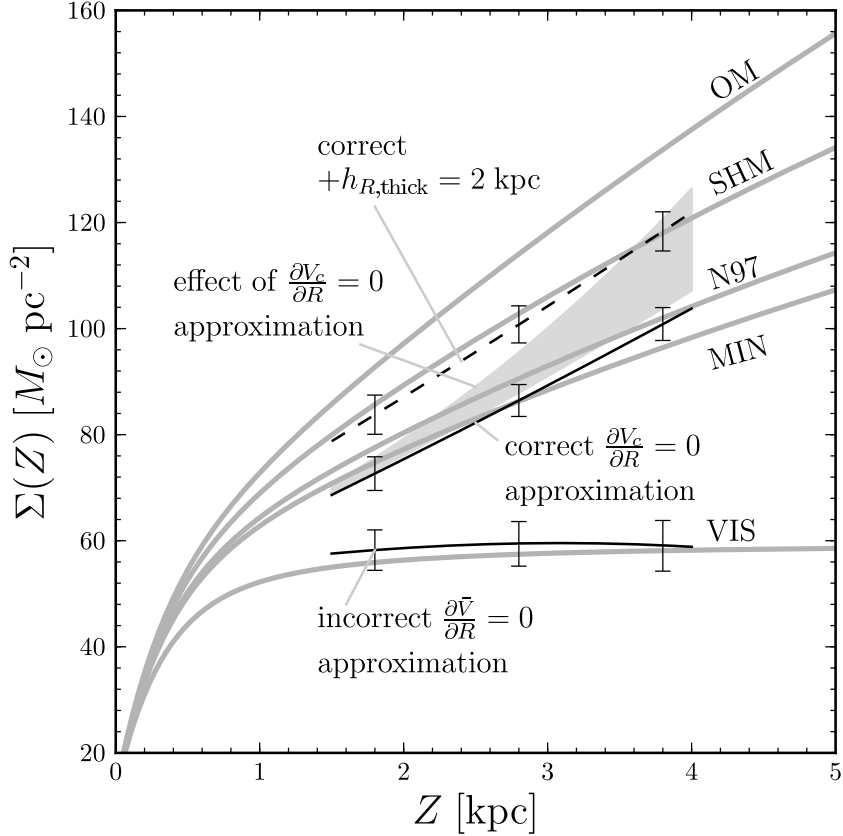}
\caption{The surface density as a function of height using the invalid
  assumption that $\partial \bar{V} / \partial R = 0$ (lower black
  curve) and the more realistic assumption that $\partial V_c /
  \partial R = 0$ in the mid-plane (upper black curve). The latter
  assumption is shown in \sectionname~\ref{sec:poisson} to give a robust lower limit to the surface
  density. The dashed curve shows
  the effect of reducing the radial scale length of the tracer from
  \mb's value $h_R=3.8\kpc$ to the more likely value of 2 kpc. Also
  shown as the gray band is the range of surface densities that
  results from applying the lower and upper curves in
  \figurename~\ref{fig:approx} to correct the approximation that $V_c$
  is independent of height; a similar gray band would apply to the
  dashed curve. 68\,\% uncertainty intervals on the observed surface
  density are shown at a few representative points. The curves
  representing estimates of the visible matter (`VIS') and the
  predictions of various dark-matter halo models (`OM',`SHM',`N97',
  and `MIN'), defined in \sectionname~\ref{sec:reanalysis}, are the
  same as in Figure 1 of \mb.}\label{fig:correct}
\end{figure}

Using the kinematics from \citet{MoniBidin12a} and \mb\ in
\eqnname~(\ref{eq:correct}), we get the curve labeled as ``correct
$\partial V_c / \partial R=0$ approximation'' in
\figurename~\ref{fig:correct}. As we have discussed, this gives a
lower limit to the surface density.  The gray shaded region in the
Figure 
shows the range of surface densities obtained after including the 
radial integral in \eqnname~(\ref{eq:poisson}), using the bottom
and top curves in \figurename~\ref{fig:approx}. These can be compared to the curve
labeled ``incorrect $\partial \bar{V} / \partial R = 0$
approximation'', which is the curve presented in \mb\ as their primary
result (their Figure 1). 

We can also compare these curves with Galactic mass models in the
literature, showing the same mass models as \mb.  The estimated
contribution to $\Sigma(Z)$ from baryonic matter (VIS for ``visual''
in \figurename~\ref{fig:correct}) is composed of a thin 13 $M_\odot$
pc$^{-2}$ layer of interstellar medium, plus a stellar halo and
``thick'' and ``thin'' disk components with parameters taken from
\citet{Juric08a}; in this model the normalization is chosen so that
$\Sigma_{\mathrm{disk}}(1.1\ \mathrm{kpc}) = 40 M_\odot$ pc$^{-2}$
\citep{Holmberg04a}. \figurename~\ref{fig:correct} also shows the
expected surface density when various dark-matter halo models are
added to this baryonic mass. The OM halo model has a profile
\begin{equation}\label{eq:psuedoiso}
  \rho_{\mathrm{DM}}(R_0,Z) = \rho_c\left(\frac{R_c^2}{R^2_c + R_0^2 + Z^2}\right)\,,
\end{equation}
with $R_c = 8.01\kpc$ and $\rho_c = 0.0103\ M_\odot$ pc$^{-3}$
\citep{Olling01a}, such that $\rho_{\odot,\mathrm{DM}} = 0.0084\ M_\odot$
pc$^{-3}$. The other halo models were all taken from
\citet{Weber10a} and have densities following
\begin{equation}\label{eq:nfwlike}
  \rho_{\mathrm{DM}}(r) = \rho_{\odot,DM}\,\left(\frac{r}{R_0}\right)^{-\alpha}\,\left(\frac{1 + (r/R_c)^\beta}{1+(R_0/R_c)^\beta}\right)^{-\gamma}\,,
\end{equation}
where $r$ is the Galactocentric spherical radius. The Standard Halo
Model (SHM) has an NFW profile ($\alpha = \beta = 1$, $\gamma = 2$)
with $R_c = 10.8\kpc$ and $\rho_{\odot,\mathrm{DM}} = 0.0084\ M_\odot$
pc$^{-3}$. The N97 model also has an NFW profile with $R_c = 20\kpc$
and $\rho_{\odot,\mathrm{DM}} = 0.0061\ M_\odot$ pc$^{-3}$. The MIN
model has a profile as in \eqnname~(\ref{eq:psuedoiso}) (i.e.,
$\alpha=0, \beta=2, \gamma=1$) with $R_c = 5\kpc$ and $\rho_{c} =
0.019\ M_\odot$ pc$^{-3}$ \citep{Weber10a}, such that
$\rho_{\odot,\mathrm{DM}} = 0.0053\ M_\odot$ pc$^{-3}$. All of these
models were constrained by assuming that $V_c(R_0) = 244 \pm 10\kms$,
except for the OM model which has $V_c(R_0) = 220\kms$. The difference
between the predictions of these models for $\Sigma(Z)$ in
\figurename~\ref{fig:correct} is mainly due to their different value
for $\rho_{\odot,\mathrm{DM}}$.

It is clear from \figurename~\ref{fig:correct} that the \mb\ data are
fully consistent with the predictions from several standard dark
matter models when using the correct assumption $\partial V_c /
\partial R = 0$. In particular, the values of the surface density
throughout the range $|Z|=1$--$4\kpc$ are consistent with the standard
halo model (SHM) which has $\rho_{\mathrm{DM}} \simeq
0.01\msun\pc^{-3}=0.38$ GeV cm$^{-3}$. \figurename~\ref{fig:correct}
also shows the effect of using the more appropriate radial scale
length $h_R = 2\kpc$, which increases the surface density even
further. The slope of the measured $\Sigma(Z)$ implies a minimum
dark-matter density of $\rho_{\mathrm{DM}} =$ 0.007$\pm$0.001
$M_\odot$ pc$^{-3} = 0.27$$\pm$0.04 GeV cm$^{-3}$ for $h_R = 3.8\kpc$,
and $\rho_{\mathrm{DM}} =$ 0.0085$\pm$0.0015 $M_\odot$ pc$^{-3} =
0.32$$\pm$0.06 GeV cm$^{-3}$ for $h_R = 2.0$ kpc. These uncertainties
are statistical and do not include the systematic uncertainty
associated with the approximation that the circular-velocity curve is
flat at all heights, which adds at most 0.1 GeV cm$^{-3}$ (see
\figurename~\ref{fig:approx}), or that associated with the measurement
of the dispersions (see, \eg, the erratic behavior of $\sigma^2_{UW}$
in figure 8 of \citealt{MoniBidin12a}), or the assumption that the
tracers follow an exponential distribution in radius and height with
the assumed scale lengths. These systematic uncertainties are at least
of the same magnitude as the statistical uncertainties. For example,
if instead of $h_Z = 900$ pc we use $h_Z = 700 $ pc, which is the
mass-weighted mean scale height in the range 2 to 4 kpc using the
measurements of \citet{Bovy12a,Bovy12b}, $\rho_{\mathrm{DM}}$
increases by 0.001 $M_\odot$ pc$^{-3}$; using a varying mass-weighted
$h_Z$ as a function of $|Z|$ as given by those same measurements
changes $\rho_{\mathrm{DM}}$ by 0.002 $M_\odot$ pc$^{-3}$. Taking
these systematics into account, our best estimate for the local
dark-halo density is
$\rho_{\mathrm{DM}}=0.008\pm0.003\msun\pc^{-3}=0.3\pm0.1\,\mbox{GeV
  cm}^{-3}$.

\section{Conclusions}\label{sec:conclusion}

In this paper, we have shown that the assumption of a radially
constant mean azimuthal velocity for the stellar tracers used by
\mb\ is physically implausible. This assumption is the reason why the
\mb\ analysis finds a constant $\Sigma(Z)$ at $2 \lesssim |Z| \lesssim
4$ kpc, in seeming contradiction with the standard expectation of a
dark-matter density $\rho_{DM} \approx 0.01\, M_\odot$ pc$^{-3} =
0.38$ GeV cm$^{-3}$ in this region. The mean azimuthal velocity
$\bar{V}$ of the warm tracer population of stars used by \mb---stars
that are part of the thicker components of the Milky Way disk that reach large
heights above and below the plane---is significantly different from
the circular velocity $V_c$, with $\bar{V} \lesssim 0.5V_c$ at $|Z| =
3\kpc$. We have shown in \sectionname~\ref{sec:vphivcirc} that for a
circular-speed curve that is close to flat, the expected $\partial
\bar{V} / \partial R$ reaches tens of $\kms\kpc^{-1}$ at a few
kpc from the Galactic mid-plane. Indeed, the assumption that $\partial
\bar{V} / \partial R = 0$ requires that the circular-speed curve falls
off in a Keplerian manner at a few kpc above the plane, in clear
contradiction with observations.

We derived an alternative formula for $\Sigma(Z)$ that assumes that
the circular-velocity curve is flat at all heights above the plane and
we showed that this approximation leads to a lower limit for all
plausible mass distributions. This approximation sidesteps the issue
of the unknown radial trend of $\bar{V}$ and as such makes no
assumptions about it. Applying this alternative formula,
we find that the \mb\ data give a lower limit that is fully
consistent with the standard local density of dark matter of
$\rho_{DM} \approx 0.01\, M_\odot$ pc$^{-3}$, and that they imply a
local dark-matter density of $0.008\pm0.003\msun\pc^{-3}=
0.3\pm0.1\,\mbox{GeV cm}^{-3}$, where the error bars include both
statistical and less well-known systematic errors. Therefore, our
analysis shows that the locally measured density of dark matter is
consistent with that extrapolated from halo models constrained at
Galactocentric distances $r \gtrsim 20 \kpc$: for example,
extrapolating the best-fit halo profiles of \citet{Xue08a} and
\citet{Deason12a}, obtained from fitting halo stars with $20 \lesssim
r \lesssim 60\kpc$, gives $\rho_{DM} \approx 0.006$ to $0.011\ M_\odot$
pc$^{-3}$.

The breakdown of the assumptions made in this simple,
``model-independent'' Jeans analysis are such that the measurement has
a systematic uncertainty reaching 10 to 20\,\% at $|Z|
=4\kpc$. Therefore, a precise determination of the local dark matter
density from observations at large $Z$ using the Jeans analysis of
\mb\ requires data that span a wide range in $R$ such that the radial
gradient of the velocity moments, foremost $\bar{V}$, can be
determined. The \emph{Gaia} mission \citep{Perryman01a} will provide
such measurements in the near future. Currently, data that span
multiple kpc in $R$ and $Z$ are available from the SDSS/SEGUE
survey \citep{Yanny09a}. In contrast to the data from
\citet{MoniBidin12a}, the SEGUE volume selection is known in detail
\citep[e.g.,][]{Bovy12b} so both the spatial structure and kinematics
can be obtained for the same set of tracer stars. The SEGUE data also
have the advantage that they contain information on the
elemental abundances of each star, allowing the tracers to be divided
into sets of stars with simple distribution functions
\citep[e.g.,][]{Bovy12b,Bovy12c}.

Finally, we note that our estimate is of the mean halo density between
1 and 4 kpc above the Galactic mid-plane. The halo density at the Sun,
which is the relevant quantity for direct dark-matter detection
experiments, is likely to be larger because of two effects. The
density in the mid-plane for a spherical NFW halo with a scale radius
of $22.25\kpc$ \citep{Xue08a} is $7\%$ larger than at $|Z| =
2.5\kpc$. Besides this purely geometric effect, the gravitational
influence of the disk further increases the mid-plane dark-matter
density. An isothermal halo with isotropic velocity dispersion
$\sigma$ has a density $\rho \propto \int \dd \vec{v}\,
e^{-E/\sigma^2} = e^{-\Phi/\sigma^2}$, so we expect that
\begin{equation}
\frac{\rho_{DM}(|Z|)}{\rho_{DM}(0)} - 1 \simeq -\frac{2\pi G\Sigma_{\rm disk}|Z|}{\sigma^2}=
-0.20\,\frac{\Sigma_{\rm
    disk}}{50\msun\pc^{-2}}\left(\frac{130\kms}{\sigma}\right)^2\frac{|Z|}{2.5\kpc}\,.
\end{equation}
These two effects imply that the dark-matter density in the mid-plane
is enhanced over the value derived in this paper by about 30\%. This
agrees with the estimated enhancement in an $N$-body simulation by
\citet{Garbari11a}.


\acknowledgements It is a pleasure to thank the anonymous referee,
Dana Casetti-Dinescu, Christian Moni Bidin, Jerry Ostriker, Justin
Read, Hans-Walter Rix, Martin Smith, and Nadia Zakamska for helpful
comments. Support for Program number HST-HF-51285.01-A was provided by
NASA through a Hubble Fellowship grant from the Space Telescope
Science Institute, which is operated by the Association of
Universities for Research in Astronomy, Incorporated, under NASA
contract NAS5-26555.

\clearpage

\clearpage

\end{document}